\documentclass[preprint,prd]{revtex4} 
 
\begin{document} 
\input{epsf}

\title{Stochastic Spacetime and Brownian Motion of Test Particles}

\author{ L.H. Ford} 
 \email[Email: ]{ford@cosmos.phy.tufts.edu} 
 \affiliation{Institute of Cosmology  \\
Department of Physics and Astronomy\\ 
         Tufts University, Medford, MA 02155}

\begin{abstract} 
The operational meaning of spacetime fluctuations is discussed. Classical
spacetime geometry can be viewed as encoding the relations between the
motions of test particles in the geometry. By analogy, quantum fluctuations
of spacetime geometry can be interpreted in terms of the fluctuations of
these motions. Thus one can give meaning to spacetime fluctuations
in terms of observables which describe the Brownian motion of test particles.
We will first discuss some electromagnetic analogies, where quantum
fluctuations of the electromagnetic field induce Brownian motion of test
particles. We next discuss several explicit examples of Brownian motion
caused by a fluctuating gravitational field. These examples include
lightcone fluctuations, variations in the flight times of photons
through the fluctuating geometry, and fluctuations in the expansion parameter
given by a Langevin version of the Raychaudhuri equation. The fluctuations
in this parameter lead to variations in the luminosity of sources. Other
phenomena which can be linked to spacetime fluctuations are spectral
line broadening and angular blurring of distant sources.
\end{abstract}

\maketitle 
 
\baselineskip=13pt 

\section{Introduction}

It is well known that classical spacetime geometry can be mapped by the motion
of classical test particles which move along geodesics in the given geometry.
Here test particle will be understood to include both massive particles,
which move on timelike geodesics, and light rays moving on null geodesics.
The fundamental quantity needed to characterize a spacetime geometry is the
Riemann tensor, which in turn can be characterized by the phenomenon of 
geodesic deviation.

A basic feature of a theory which combines quantum theory and gravitation
is expected to be fluctuations of the spacetime geometry. These
fluctuations can arise either from the quantum nature of the gravitational 
field itself (active fluctuations), or from quantum fluctuations of the
matter stress tensor\cite{F82,Kuo,PH97} (passive fluctuations). In general, 
both types of fluctuations can be present 
simultaneously~\cite{CH95,CCV,MV99,HS98}. 

The key question which we wish to address is how to give an operational
significance to fluctuations of the spacetime geometry. Test particles will
no longer follow fixed classical geodesics, but will rather undergo
Brownian motion around a mean geodesic.This Brownian motion can be
described by the quadratic fluctuations of some quantity characterizing
the variation from the mean geodesics. Some explicit examples will be
treated below.

\section{Electromagnetic Analogies}

In this section, we will discuss some examples of fluctuating forces of
electromagnetic origin which are useful as analogies to the effects of the
fluctuations of gravity. Specifically, we will examine the Brownian motion 
of a charged test particle coupled to a fluctuating electromagnetic field,
radiation pressure fluctuations of a mirror, and Casimir force fluctuations.

\subsection {Classical Brownian Motion}

Consider a nonrelativistic charged particle of mass $m$ and charge $q$
in the presence of an electric field ${\mathbf E}$. If we ignore magnetic
forces, the particle's velocity satisfies
\begin{equation}
\frac{d {\mathbf v}}{d t} = \frac{q}{m}\, {\mathbf{E}}({\mathbf{x}},t)\,.
                                  \label{eq:lang0}
\end{equation}
If the particle starts at rest at time $t=0$, then we can write
\begin{equation}
 {\mathbf{v}}={q\over m}\int_0^t\;{\mathbf{E}}({\mathbf{x}},t)\;dt \, ,
                                  \label{eq:v}
\end{equation}
where in general $\mathbf{x}=\mathbf{x}(t)$. Here we will assume that the
particle does not move very far on the time scales of interest, so we can
ignore the time dependence of $\mathbf{x}$. 

Now suppose that the electric field undergoes fluctuations, so the mean
trajectory of the particle is obtained by averaging Eq.~(\ref{eq:v}):
\begin{equation}
 \langle{\mathbf{v}}\rangle =
{q \over m}\int_0^t\;\langle{\mathbf{E}}({\mathbf{x}},t)\rangle \;dt \, .
                                  \label{eq:v_av}
\end{equation} 
The fluctuations around the mean trajectory in the $i$-direction
are described by
\begin{equation}
\langle{\Delta v_i^2}\rangle={q^2\over m^2}\;\int_0^t\;\int_0^t\;
[\langle{E}_i({\mathbf x},t_1)\;{E}_i({\mathbf x},t_2)\rangle
-\langle{E}_i({\mathbf x},t_1)\rangle\;
    \langle{E}_i({\mathbf x},t_2)\rangle]\;dt_1\;dt_2\;.\label{eq:lang}
\end{equation}
Thus, the dispersion in the $i$-component of velocity is given as a
double time integral of an electric field correlation function.
Let $C$ denote this correlation function, and assume that it is a function
of the time difference $\tau = |t_1 -t_2|$ only, so we can write
\begin{equation}
C(\tau) = \langle{E}_i({\mathbf x},t_1)\;{E}_i({\mathbf x},t_2)\rangle
-\langle{E}_i({\mathbf x},t_1)\rangle\;
    \langle{E}_i({\mathbf x},t_2)\rangle\,.
\end{equation}
In this case, we may write Eq.~(\ref{eq:lang}) as
\begin{equation}
\langle{\Delta v_i^2}\rangle=2 \,{q^2\over m^2}\;\int_0^t\;(t -\tau)\,
C(\tau)\, d\tau \, .  \label{eq:lang2}
\end{equation}

\begin{figure}
\begin{center}
\leavevmode\epsfysize=6cm\epsffile{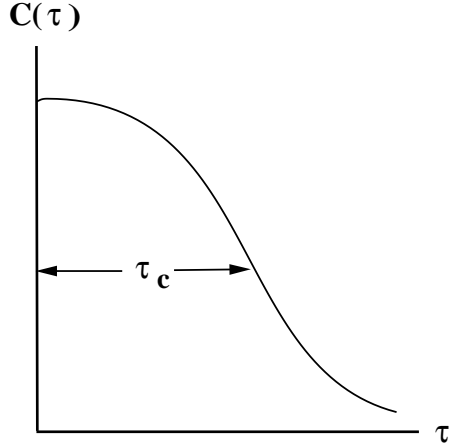}
\end{center}
\caption{A correlation function typical of classical fluctuations. 
The correlation function $C(\tau)$ is non-negative, and decreases 
monotonically. The characteristic width of $C(\tau)$ is the correlation time, 
$\tau_c$. }
\label{fig:corr_cla }
\end{figure}

The simplest possibility is classical or thermal-like fluctuations.
Here the correlation function 
$C$ is non-negative, 
and decays monotonically on a timescale of $\tau_c$, the correlation time,
as illustrated in Fig.~\ref{fig:corr_cla }.
 In this case, performing the integration in 
Eq.~(\ref{eq:lang2}) will yield a result of the form
\begin{equation}
\langle{\Delta v_i^2}\rangle = a\,{q^2\over m^2}\; C(0)\, \tau_c \;t \, ,
\end{equation}
where $a$ is a constant of order one. This is the familiar random walk
behavior where $v_{\rm rms} = \sqrt{\langle{\Delta v_i^2}\rangle}
\propto \sqrt{t}$. Note that here we are ignoring any damping effects,
which will eventually stop the growth of  $v_{\rm rms}$.

\subsection{Quantum Fluctuations}

Now we wish to consider a model in which quantum, as opposed to classical
fluctuation, drive the Brownian motion of the test particle. It is not clear
that there is any effect associated with fluctuations of the quantized
electromagnetic field in empty flat spacetime. However, there is a nontrivial
effect produced by {\it changes} in these fluctuations, as will occur
in the presence of a boundary. The simplest such boundary is a perfectly
reflecting plane, and the resulting effects on Brownian motion of a charged
particle were recently discussed in Ref.~\cite{YF04}. Here we will summarize
the results of that paper. Let the reflecting plane be at $z = 0$, so we 
need to consider motion of the test particle in both the longitudinal ($z$)
direction, and in a transerse ($x$) direction. 

\begin{figure}
\begin{center}
\leavevmode\epsfysize=8cm\epsffile{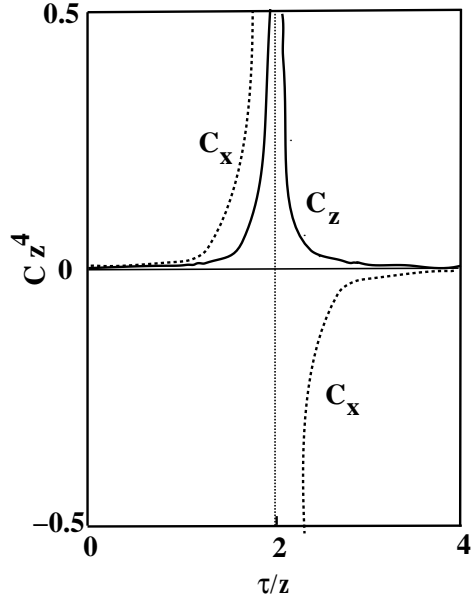}
\end{center}
\caption{The renormalized electric field correlation functions near a 
perfectly reflecting mirror are plotted as functions of the time separation
$\tau$ for fixed distance $z$ from the mirror. The solid line gives the
correlation function for the longitudinal components, $C_z$, and
the dotted line that for the transverse component, $C_x$. The dimensionless
quantities  $C_z\, z^4$ and $C_x\, z^4$ are plotted.  Both functions
are singular at $\tau = 2 \, z$. }
\label{fig:corr_qua}
\end{figure}

The appropriate correlation functions are
the renormalized two-point functions, which are the differences in
the two-point functions with and without the plate,
\begin{equation}
C_z(\tau,z) = \langle{
 E}_z({\mathbf x},t')\;{E}_z({\mathbf
 x},t'')\rangle={ 1 \over \pi^2 ( \tau^2 -
 4z^2)^2}\;,
\end{equation}  
and
\begin{equation}
C_x(\tau,z) =\langle{ E}_x({\mathbf x},t')\;{ E}_x({\mathbf
 x},t'')\rangle=\langle{E}_y({\mathbf x},t')\;{E}_y({\mathbf
 x},t'')\rangle=-{ \tau^2 + 4z^2\over \pi^2 ( \tau^2 -
 4z^2)^3} \, .
\end{equation}
Note that here $\langle {\mathbf E} \rangle = 0$. These correlation functions
are plotted in Fig.~\ref{fig:corr_qua}. 
They are singular at $\tau = 2\,z$, corresponding to the roundtrip light 
travel time to the plate, and can be either positive or negative.

Despite the singularity, which can be attributed to
the use of perfectly reflecting boundary conditions and a plate with a sharp
boundary, the integral in
Eq.~(\ref{eq:lang2}) can be performed by an integration by parts approach.
The results, for $t \gg z$ are
\begin{equation}
\langle \Delta v_x^2\rangle \approx\;-{q^2\over 3\pi^2
m^2}{1\over t^2}-{8q^2\over 5 \pi^2 m^2}{z^2\over t^4}\;,  \label{eq:vx2}
\end{equation}
and
\begin{equation}
 \langle \Delta v_z^2\rangle\approx\;{q^2\over 4 \pi^2 m^2}{1\over z^2}
+ {q^2\over 3\pi^2 m^2}{1\over t^2}  \;. \label{eq:vz2}
\end{equation}
Unlike the case of classical noise, here the mean squared velocities
do not grow in time. This is required by energy conservation, as there is no
source of energy from which the particles can acquire kinetic energy. The
fact that $\langle \Delta v_x^2\rangle \not= 0$ reflects the fact that some 
energy is required to set up the system at $t=0$, that is, a transient effect.
The mathematical reason that we do not find growing mean squared velocities
is that the correlation functions have both positive and negative regions
which tend to cancel one another so that
\begin{equation}
\int_0^\infty C_x(\tau,z)\, d\tau = \int_0^\infty C_z(\tau,z)\, d\tau = 0\,.
\end{equation}
Unlike a thermal state, a quantum state such as we are considering here
is highly correlated, with subtle correlations and anticorrelations.

It may come as a surprize that that the integral of $C_z$ vanishes, as
this appears to be a positive function. However, the singularity at
$\tau = 2z$ effectively contributes a negative contribution when the
integral is defined by integration by parts. A simple example of this 
is the following:
\begin{equation}
\int_{-\infty}^\infty \frac{d x}{x^2} = - \int_{-\infty}^\infty dx \;
\frac{d}{d x}\, \left(\frac{1}{x}\right) = 0 \,.
\end{equation}
The apparently positive function $1/x^2$, when defined as a distribution,
has a negative part at $x=0$. We might be able to remove
the singularities in $C_x$ and $C_z$ by a more realistic model of the
plate, such one which includes dispersion and surface roughness. 
However, if this were done, then $C_z(\tau)$ would become a finite function
with a negative region near $\tau = 2 z$. Similar behavior was found for
the mean squared electric field near a plate when the singularity was smeared 
out by position fluctuations~\cite{FS98}.

The mean squared displacements in both the transverse and longitudinal
directions can be found by integrating the Langevin equation, 
Eq.~(\ref{eq:lang0}) twice and then squaring. In  Ref.~\cite{YF04}, it
is shown that the mean squared position fluctuations are, in the  $t \gg z$
limit,
\begin{equation}
\langle \Delta x^2\rangle \approx\;-{q^2\over
3\pi^2m^2}\ln(t/2z)\;,
\end{equation}
and
\begin{equation}
 \langle \Delta z^2\rangle \approx \frac{q^2}{\pi^2 m^2}\, \left[
\frac{t^2}{8 z^2} + \frac{1}{3} \ln\left(\frac{t}{2z}\right) + \frac{1}{9}
+ O(z^2/t^2) \right] \;.  \label{eq:z2_large}
\end{equation}
It is of particular interest to note that $\langle \Delta x^2\rangle < 0$.
This can only be understood if we account for the quantum nature of the
test particles, and  amounts to a reduction is the usual quantum position
uncertainty of the particle. The results summarized in this subsection
can also be generalized to the case of two parallel plates~\cite{YC04}.

\subsection{ Fluctuations of Radiation Pressure and Casimir Forces}

The model discussed in the previous subsection can be viewed as an
analog model for active metric fluctuations in that the charged particle is
coupled directly to the quantized electric field. Here we wish to discuss
some situations which are analogous to the passive metric fluctuations.
These arise when quantum fluctuations of the electromagnetic field stress 
tensor cause force fluctuations on uncharged material bodies. 

One example of this is the fluctuations in radiation pressure when light
in a coherent state is reflected by a mirror. 
This effect was first discussed by
Caves~\cite{Caves1,Caves2} using an approach based upon fluctuations in the
numbers pf photons striking the mirror. Caves' results were later rederived 
and extended using an approach~\cite{WF01} based upon the quantum stress 
tensor. This approach requires an integration of a state-dependent, but 
singular cross term, which must be performed by an integration by parts
procedure. Consider a free mirror of mass $m$ and area $A$ on which 
a linearly polarized beam of light in a coherent state with energy density 
$\rho$ shines perpendicularly for a time $t$. The resulting uncertainty
in the mirror's velocity is 
\begin{equation}
\label{eq:dv2_{s}t}
\langle \triangle v^{2}\rangle =4\, \frac{A\omega \rho }{m^{2}}\, 
t \, .
\end{equation}
This uncertainty is likely to be a significant source of noise in
future generations of laser interferometry detectors of gravity waves.
When it is detected experimentally, it will constitute an observation
of quantum stress tensor fluctuations.  

Another phenomenon due to electromagnetic stress tensor fluctuations 
are the fluctuations of Casimir forces, which have been discussed by
several authors~\cite{Barton,Eberlein,JR,WKF01}. An example is the
fluctuation of the Casimir-Polder force on an atom in the presence
of a reflecting plate discussed in Ref.~\cite{WKF01}. Consider an atom of
mass $m$ and static polarizability $\alpha$ at a distance $z$ from
the plate. The transverse velocity fluctuations are found to be
\begin{equation}
   \langle\triangle v_{x}^{2}\rangle
   =\frac{47}{768}\frac{\hbar^2 \,\alpha^{2}}{\pi^{4}m^{2}z^{8}}
    \label{eq:final_x} 
\end{equation}
and 
\begin{equation}
   \langle\triangle v_{z}^{2}\rangle
   =-\frac{3787}{3840}\frac{\hbar^2 \,\alpha^{2}}{\pi^{4}m^{2}z^{8}}
    \label{eq:final_z} \, .
\end{equation} 
These results are analogous to Eqs.~(\ref{eq:vx2}) and (\ref{eq:vz2}) 
for the charged
particle case. Again we can find {\it suppression} of the usual
quantum uncertainty. In both cases, there are subtle correlations
in the quantum fluctuations, which is a typical feature of quantum
stress tensor fluctuations~\cite{FR04}.

\section{Fluctuating Spacetime}

Now we turn to a discussion of a few of the phenomena which can arise
when spacetime geometry fluctuates.

\subsection{A Loophole in the Classical Singularity Theorems}

The singularity theorems proven by Penrose, Hawking and others show
that formation of a singularity is required in classical general
relativity provided that certain energy conditions are obeyed. It
has long been recognized that quantum matter fields can violate these
classical energy conditions, for example in quantum states with 
negative local energy densities. This allows the possibility of
singularity avoidance in semiclassical gravity, where a classical
gravitational field is coupled to the expectation value of a quantum
matter stress tensor.

It is less well known that even very small fluctuations in spacetime
also create a distinct loophole in the singularity theorems~\cite{F03}. 
The reason for
this is that the proofs of the theorems all rely upon focusing arguments.
However, a bundle of geodesics will not come to an exact focus when
there are fluctuations around a background geometry. This does not mean that
minute fluctuations will necessarily prevent the formation of a singularity,
but the classical arguments for singularity formation need to be revised.

\subsection{Lightcone Fluctuations}

One particularly striking effect of spacetime geometry fluctuations
is that there is no longer a fixed lightcone. Recall that the lightcone
plays a central role in classical relativity theory, as the boundary between
events which can be causally connected (timelike or null separations)
and those which cannot (spacelike separations). Once the  spacetime geometry
undergoes fluctuation, no matter how small, this strict distinction
can no longer be maintained. 

Here we will summarize an approach developed in Refs.~\cite{F95,FS96,FS97,
YF99}. (For other related discussions, see Ref~\cite{AC98,AC99,NV00}.)
Let us consider a nearly flat spacetime with small metric fluctuations,
which can be described by a correlation function 
$\langle h_{\mu\nu}(x) h_{\rho\lambda}(x') \rangle$, where we assume
that $\langle h_{\mu\nu}(x) \rangle = 0$. The metric fluctuations
could be due either to gravitons in a nonclassical state (active
fluctuations), or to fluctuations of a quantum matter stress tensor
(passive fluctuations).
 Consider a source
located at $r=r_0$ and a detector at  $r=r_1$. Then the mean flight
time of light rays between the source and the detector is the classical
result, $\Delta r = r_1 - r_0$. However, there will be a root-mean-squared
variation around this mean value of $\Delta t$, given by
\begin{equation}
\Delta t = \frac{\sqrt{\langle \sigma_1^2 \rangle}}{\Delta r}\, ,
\end{equation}
where
\begin{equation}
\langle \sigma_1^2 \rangle ={1\over 8}(\Delta r)^2
\int_{r_0}^{r_1} dr \int_{r_0}^{r_1} dr'
\:\,  n^{\mu} n^{\nu} n^{\rho} n^{\lambda}
\:\, \langle h_{\mu\nu}(x) h_{\rho\lambda}(x') \rangle \,.
\end{equation}
Here $n^\mu$ is the tangent vector to the mean null geodesic separating
the source and detector, and $\sigma_1$ is the first order shift in
the invariant interval $\sigma$ separating the events of emission and 
detection.

Here  $\Delta t$ represents a mean time delay or advance relative to
the classical propagation time  $\Delta r$ on the background spacetime. Thus,
if we send a sequence of pulses between the source and detector, some
will take longer than the classical time, but some will arrive sooner. 
This raises
some interesting issues concerning causality. Normally,  signals sent 
outside of the lightcone could be used to send information into the past.
On a flat background, the order of emission and detection of a pulse
traveling on a spacelike path can be changed by a Lorentz transformation.
In the case of spacetime fluctuations produced by a bath of gravitons
or by quantum matter fields in a nonvacuum state, there do not seem to be any
causal paradoxes. The source of the fluctuations defines a preferred frame 
of reference, and effectively breaks Lorentz invariance. The case of
lightcone fluctuations in the vacuum on a Minkowski spacetime is more subtle,
and will probably require a more complete quantum theory of gravity to be
understood.

\subsection{Fluctuations of the Expansion: the Langevin-Raychaudhuri Equation}

The tendency of a gravitational field to act like a lens and to focus a
bundle of geodesics is described by the Raychaudhuri equation,
Consider a congruence of either timelike or null geodesics with affine
parameter $\lambda$ and tangent vector field $k^\mu$. The  Raychaudhuri 
equation gives the rate of change of  the expansion $\theta$ along the 
congruence to be~\cite{Wald}
\begin{equation}
\frac{d \theta}{d \lambda} = - R_{\mu\nu} k^\mu k^\nu - a\, \theta^2
-\sigma_{\mu\nu} \sigma^{\mu\nu} + \omega_{\mu\nu}  \omega^{\mu\nu} \,.
                                                 \label{eq:ray}
\end{equation}
Here $R_{\mu\nu}$ is the Ricci tensor, $\sigma^{\mu\nu}$ is the shear 
and $\omega^{\mu\nu}$ is the vorticity
of the congruence. The constant $a=1/2$ for null geodesics, and $a=1/3$ for 
timelike geodesics. The expansion parameter $\theta$ is the logarithmic 
derivative of the cross sectional area $A$ of the bundle:
\begin{equation}
\theta = \frac{d\, \log A}{d \lambda} \,,
\end{equation}
so that $\theta < 0$ describes the case of converging geodesics.
For ordinary matter obeying classical energy conditions, the Ricci tensor
term acts to decrease $\theta$, corresponding to the tendency of gravity
to focus light rays. The Raychaudhuri equation plays a crucial role in
the proofs of the classical singularity theorems.

 We are interested in interpreting this equation as a 
Langevin equation in which the Ricci tensor fluctuates. A more detailed 
account of this work is given in Ref.~\cite{BF04}.
For the purposes of 
this paper, we will further assume that the shear and vorticity of the 
congruence vanishes, and that the expansion remains sufficiently small that
the $\theta^2$ term can also be ignored, so we can write
\begin{equation}
\frac{d \theta}{d \lambda}  = - R_{\mu\nu} k^\mu k^\nu \, ,
\end{equation}
The variance of the expansion can be expressed as a double integral of the
Ricci tensor correlation function as
\begin{equation}
\langle (\Delta \theta)^2 \rangle=
\langle \theta^2 \rangle- \langle \theta \rangle^2 =
\int_0^{\lambda_0} d\lambda \int_0^{\lambda_0} d\lambda' \;
 C_{\mu \nu \alpha \beta}(\lambda,\lambda')\,\; k^\mu (\lambda) k^\nu (\lambda)
\; k^\alpha(\lambda') k^\beta(\lambda') \,, \label{eq:var}
\end{equation}
where
\begin{equation}
C_{\mu \nu \alpha \beta}(x,x') = 
\langle R_{\mu \nu }(x) R_{\alpha \beta}(x') \rangle
- \langle R_{\mu \nu }(x) \rangle \langle R_{\alpha \beta}(x') \rangle \,.
\end{equation}

Expansion fluctuations are sensitive only to the Ricci part of the
curvature, and hence to the passive fluctuations induced by the stress tensor
fluctuations. Thus when both active and passive fluctuations are present,
the expansion fluctuations provide a signature of the passive part.

As discussed in Ref.~\cite{BF04}, the Ricci tensor correlation function has
both state-independent and state-dependent singularities when $x$
and $x'$ are null separated. These singularities can removed by
replacing the double line integration in Eq.~(\ref{eq:var}) by integrations
over a four-dimensional worldtube, corresponding to the spacetime
volume occupied by a finite wavepacket. One then needs an integration
by parts procedure to define the resulting four-dimensional
integrals.

Next we need a physical interpretation of the result for $\Delta \theta$.
One such interpretation is a description of luminosity fluctuations.
The focusing of a bundle of rays, where gravity acts as a converging
lens, increases the apparent luminosity of a source. Conversely, 
defocusing (a decrease in $\theta$) decreases the apparent luminosity.
Thus fluctuations in $\theta$ produce variations in the observed 
luminosity of a source. This is a well-known phenomenon in astronomy.
Density fluctuations in the Earth's atmosphere lead to the twinkling of
stars ("scintillation" in astronomer's terminology). In principle,
quantum fluctuations of spacetime geometry could also produce
scintillation of distant sources. 

In general, the fractional variation in luminosity of a source,
$\Delta L/L$, is given by a double integral of a $\theta$-$\theta$
correlation function:
\begin{equation}
\left\langle\;\left(\Delta L \over L\right)^{2}\;\right\rangle = \int_{0}^{s} 
 \int_{0}^{s}dt'\; dt'' \;\; \left[ \left\langle\; \theta(t')\;\theta(t'')\;
\right\rangle -\langle \theta(t')\rangle\;\langle\theta(t'')\rangle \right]  .
\end{equation}
However, in some cases such as classical noise, the fractional luminosity 
fluctuations are
directly related to the value of $\Delta \theta$,
\begin{equation}
\left(\Delta L \over L\right)_{\rm rms} \approx s\, \Delta \theta \,,
\end{equation}
where $s$ is the flight distance.

There is a distinct physical effect which may be estimated from a
calculation of $\Delta \theta$. This is the degree of angular blurring 
of a distant source due to passive metric fluctuations. In Ref.~\cite{BF04} ,
an argument is given which relates the root-mean-squared variation
in angular position, $\Delta \varphi$, of a source to the variance of
$\theta$, with the result
\begin{equation}
\Delta \varphi_{\rm rms} = \frac{1}{2}\, s \, \Delta \theta_{\rm rms} \,.
                                            \label{eq:delphi}
\end{equation}
This result is just a heuristic estimate, and a more precise formula
for $\Delta \varphi$ will be derived in the next subsection. For another
discussion of the propagation of light in a fluctuating geometry, see
Hu and Shiokawa~\cite{HS98}.

In the Minkowski vacuum state, the result obtained from Eq.~(\ref{eq:var})
is formally infinite, unless we average over a finite bundle of geodesics.
Let this averaging be described by two length scales, $a$ and $b$. 
Here $a$ is the characteristic time during which the bundle of photons is
emitted, and $b$ is the typical cross sectional dimension of the bundle,
that is, its size in a spatial direction perpendicular to the direction of
propagation. The result is found to be of the form
\begin{equation}
\langle (\Delta \theta)^2 \rangle_{vac} = \ell_P^4\,
\frac{256\, A_8\, a^2}{5 \pi^2 b^8} \,, \label{eq:vac_theta}
\end{equation}
where $\ell_P$ is the Planck length, and $A_8$ is expected to be of order one.
This leads to the estimates
\begin{equation}
2 \Delta \varphi = 
\left({\Delta L\over L}\right)_{rms} = 0.1 \,A_8\, \ell_P^2 \, \frac{a\, s}{b^4} \approx
10^{-8}\,A_8\, \left(\frac{a}{b}\right)\, 
\left(\frac{10^{-10} {\rm cm}}{b}\right)^3 \,
\left(\frac{s}{10^{28} {\rm cm}}\right) \, .   \label{eq:vac_phi}
\end{equation}
Even if we were to take  $a$ and $b$ to be of the order of the photon 
wavelength, the smallest values they could reasonably have, this
effect is too small to observe. It is plausible that one should never
observe the effects of stress tensor fluctuations in the Minkowski vacuum,
but that they should be masked by the usual quantum uncertainty of the
test particles.

A nonvacuum state, such as a thermal state, is a different story. In
the case of a thermal bath at temperature $T$ and a bundle of rays
which are localized on a scale small compared to $1/T$, one finds
\begin{equation}
\Delta \theta_{rms}= \frac{128 \sqrt{c_0}}{\pi} \, \ell_P^2 \, 
\sqrt{s\, T^7} \, ,
                                                \label{eq:del_theta}
\end{equation} 
where $s$ is the flight distance and $c_0 \approx 0.3468$. Here
we find the $\sqrt{s}$ behavior characteristic of a random walk.
This leads to the estimate
\begin{equation}
\left({\Delta L \over L}\right)_{rms} = 0.02\, 
\left(\frac{s}{10^{28} {\rm cm}} \right)^\frac{3}{2}
\, \left(\frac{T}{10^6 K} \right)^\frac{7}{2}  =
10^{-3}\, \left(\frac{s}{10^{6} {\rm km}} \right)^\frac{3}{2}
\, \left(\frac{T}{1 {\rm GeV}} \right)^\frac{7}{2} \,. \label{eq:estimates}
\end{equation}
Although the effects of thermal stress tensor fluctuations are typically
small, they are in principle observable far from the Planck scale.

The possibility of observable effects of stress tensor fluctuations 
due to compact extra dimensions was discussed in Ref.~\cite{BF04b}.
The basic idea is that the smaller the compact extra dimension, the
more violent
will be the quantum stress tensor fluctuations of the Casimir energy
due to the compactification, potentially leading to observable
angular blurring or luminosity fluctuations. It is true that if we hold
the Newton's constant in the higher dimensional space fixed, one finds that
\begin{equation}
\Delta \theta_{rms} \propto \frac{G_{4+n}}{V_C}\, ,
\end{equation} 
where $G_{4+n}$ is the  Newton's constant in $4+n$ dimensions, and
$V_C$ is the volume of the $n$-dimensional compact subspace. However,
in Kaluza-Klein theories, $G_{4+n}$ is related to the Newton's constant 
in four dimensions by 
\begin{equation}
G_{4+n} = V_C\, G_{4} \,.
\end{equation}
This causes the observable effects to become independent of $V_C$,
and in fact the same as in the Minkowski vacuum in a four dimensional
spacetime. Thus one cannot detect the existence of compact extra
dimensions through angular blurring or luminosity fluctuations.

\subsection{Redshift Fluctuations and Angular Blurring}

As has been noted above, two of the physical effects of the fluctuations
of spacetime are the angular blurring of images and fluctuations in
gravitational redshifts or blueshifts. In this subsection, we will
provide a unified treatment of both phenomena in terms of the 
Riemann tensor correlation function.

\begin{figure}
\begin{center}
\leavevmode\epsfysize=8cm\epsffile{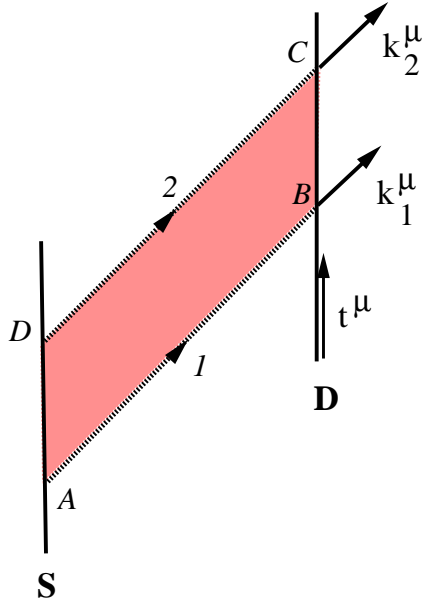}
\end{center}
\caption{A pair of null rays, {\it 1} and {\it 2} are emitted by the source
{\bf S} and detected by the detector {\bf D}. Ray {\it 1} is emitted at point 
$A$ and detected at point $B$, while ray {\it 2} is emitted at point $D$ and 
detected at $C$. The tangent vector to ray {\it 1} at $B$ is $k^\mu_1$, 
and that
to ray {\it 2} at $C$ is  $k^\mu_2$. The four-velocity of the detector is
$t^\mu$. Both the source and detector are moving along geodesic paths.
The shift in the null vector  $k^\mu$ is obtained as an integral of
the Riemann tensor over the shaded region enclosed by the null and timelike
geodesics.  }
\label{fig:s_d}
\end{figure}

First consider a source and a detector moving along geodesics in a
classical gravitational field, as illustrated in Fig.~\ref{fig:s_d}. 
Let $t^\mu$ be the four-velocity of the detector and $k^\mu$ be the
tangent vector to the null geodesics separating the events of emission
and detection. The detected frequency of photons will be proportional
to $-k^\mu\, t_\mu$. The constant of proportionality depends upon the
choice of normalization (affine parameter) for $k^\mu$.  Let {\it 1} and {\it 2} 
label two successive null geodesics, and $k^\mu_1$ and $k^\mu_2$ be 
the values of $k^\mu$ at the endpoints of each geodesic. Further
choose the normalization so that $k^\mu_1 = (1,1,0,0)$ in the frame of 
the detector. This amounts to choosing the affine parameter to
coincide with the detector's proper time at this point. Then
the fractional change in detected frequency between  {\it 1} and {\it 2} is
\begin{equation}
\left(\Delta \omega \over \omega \right) = - (k^\mu_2 - k^\mu_1)\, t_\mu \,.
\end{equation}

We are interested in the case where the frequency shift  
$\Delta \omega/ \omega$ arises from the effects of gravity, rather than from 
any change in the output of the source. This can be enforced by requiring
that the values of $k^\mu$ at the starting points of each null geodesic
be related by parallel transport along the worldline of the source.
Because  {\it 1} and {\it 2} are geodesics with tangent  $k^\mu$, the values
of  $k^\mu$ at the starting and end points of each geodesic are related
by parallel transport. Thus if we parallel transport $k^\mu_1$ backwards
along {\it 1}, along the worldline of the source to {\it 2}, 
and then along {\it 2}
to the detector, the result will be $k^\mu_2$. Recall that if we parallel
transport a vector around a closed path, the change in the vector can
be expressed as an integral of the Riemann tensor over the area enclosed 
by the path. If the detector is in a flat spacetime region, then there
is no effect from the parallel transport along the detector's worldline,
and $\Delta k^\mu = k^\mu_2 - k^\mu_1$ is the change after transport 
around the closed path. Let $t^\mu$ be defined off of the detector's
worldline as the four-velocity of a congruence of timelike geodesics
which include the worldlines of both the source and the detector. Then
we can write
\begin{equation}
\Delta k^\mu = \int {R^\mu}_{\alpha\nu\beta}\,k^\alpha\,t^\nu\,k^\beta\, 
da \, ,
\end{equation}
where the integration is over the 2-surface bounded by the four geodesic 
segments, the shaded region in Fig.~\ref{fig:s_d}. 

Now suppose that the Riemann tensor is subject to fluctuations (active
or passive or both) described by the correlation function
\begin{equation}
C_{\mu\alpha\nu\beta\; \rho\gamma\sigma\delta}(x,x') = 
\langle R_{\mu\alpha\nu\beta}(x) \, R_{\rho\gamma\sigma\delta}(x') \rangle
- \langle R_{\mu\alpha\nu\beta}(x) \rangle
 \,\langle R_{\rho\gamma\sigma\delta}(x') \rangle \,.
\end{equation}
Then we can express the variance of the fractional redshift fluctuations
as
\begin{equation}
\delta \omega^2_{\rm rms} = 
\left\langle\;\left(\Delta \omega \over \omega \right)^{2}\;\right\rangle
- \left\langle\;\Delta \omega \over \omega\;\right\rangle^2 =
\int da\, da' \; C_{\mu\alpha\nu\beta\; \rho\gamma\sigma\delta}(x,x')\;
 t^\mu k^\alpha t^\nu k^\beta\;  t^\rho k^\gamma t^\sigma k^\delta\, ,
\end{equation}
where the indices $\mu\alpha\nu\beta$ are at point $x$ and the indices
$\rho\gamma\sigma\delta$ are at $x'$.

We can also relate the degree of angular blurring to the Riemann
tensor correlation function. Let $s^\mu$ be a unit spacelike vector in 
a direction orthogonal to the direction of propagation of the null
rays; thus $s^\mu t_\mu = s^\mu k_\mu =0$. Then 
\begin{equation}
\Delta k^\mu s_\mu = \tan \varphi \approx \varphi \, ,
\end{equation}
where $\varphi$ is the change in angle between {\it 1} and {\it 2} in the plane
defined by the pair of spacelike vectors $s^\mu$ and $n^\mu = k^\mu -t^\mu$.
Here we are assuming that $|\varphi| \ll 1$. We can express $\varphi$
in terms of an integral of the Riemann tensor as
\begin{equation}
\varphi = \Delta k^\mu s_\mu = 
 \int R_{\mu\alpha\nu\beta}\,s^\mu\,k^\alpha\,t^\nu\,k^\beta\, da \,.
\end{equation}
The variance of $\varphi$ due to Riemann tensor fluctuations can be
expressed as
\begin{equation}
\Delta \varphi^2 = \langle \varphi^2 \rangle -\langle \varphi \rangle^2
= \int da\, da' \; C_{\mu\alpha\nu\beta\; \rho\gamma\sigma\delta}(x,x')\;
 s^\mu k^\alpha t^\nu k^\beta\;  s^\rho k^\gamma t^\sigma k^\delta\, .
\end{equation}
In cases where the Riemann tensor fluctuation are dominated by
Ricci tensor fluctuation, the above result can be expressed in terms of
the Ricci tensor correlation function. In general, we expect the heuristic
estimate Eq.~(\ref{eq:delphi}) to agree in order of magnitude, 
but not in detail.

\section{Summary and Discussion}

In this paper, we have discussed the idea that the operational meaning
of stochastic spacetime lies in quantities which describe the Brownian motion
of test particles, either massless or massive. Some electromagnetic analogies,
such as the motion of an electron or an atom in the modified vacuum 
fluctuations near a reflecting plate provide useful insights into the
types of effects one might expect in quantum gravity. We examined several
specific examples of Brownian motion in a fluctuating gravitational field,
These include lightcone fluctuations, whereby the flight time of pulses
between a source and a detector can fluctuate around the classical flight
time. One of the striking features of this effect is that the rigid
distinction between timelike and spacelike intervals can no longer be 
maintained when gravity fluctuates. We also reviewed recent work which
treats the Raychaudhuri equation as a Langevin equation to compute
fluctuations in the expansion of a bundle of geodesics. These fluctuations
translate into luminosity fluctuations of a source. Thus distant objects
would appear to ``twinkle'' when seen through a gravitational field
with a fluctuating Ricci tensor. Finally, we developed a formalism which
gives a unified treatment of two other effects, redshift fluctuations
and angular blurring. Both of these effects can arise when the Riemann
tensor in the region between a source and a detector undergoes fluctuations.

In most situations in the present day universe, these effects are small,
but they can in principle be large far from the Planck scale. Furthermore,
they could be significant in the early universe and other strong
gravitational field situations.

\begin{acknowledgments}
I have benefitted from discussion with numerous colleagues, including
J. Borgman, B.L. Hu, C.I. Kuo, K. Olum, T. Roman, A. Roura, N. Svaiter,
E. Verdaguer, C.H. Wu, and H. Yu.
  This work was supported in part by the National
Science Foundation under Grant PHY-0244898.
\end{acknowledgments}

\end{document}